\documentstyle{article}

\def\Evr{E90}
\def\TC{TC92}
\def\PTC{Paper~1}
\def\CPT{CPT}
\def\MPT{MPT}
\def\SPH{SPH}
\def\P3M{P$^{3}$M}

\def\APS{AdP$^{3}$M-SPH}
\def\AP3M{AdP$^{3}$M}

\def\fig{Fig.}

\def\Rmax{R_{max}}
\def\dev{de Vaucouleurs}
\def\brem{bremsstrahlung}

\def\fs{dg1.0}
\def\ff{dg2.0}
\def\fh{dg0.5}
\def\go{g1.0}
\def\acf{d1.0}
\def\AaA{A\&A}

\def\AJ{AJ}

\def\ApJ{ApJ}
\def\ApJL{ApJ}
\def\ApJS{ApJS}

\def\ARAA{ARA\&A}

\def\JCP{J.~Comp.~Phys.}
\def\MN{MNRAS}

\def\PhD{PhD thesis}
\def\Pre{Preprint}
\def\prep{in preparation}

\def\RMP{Rev.~Mod.~Phys.}

\def\etal{{\it et al.\thinspace}}

\def\ie{{\it i.e.\ }}

\def\a{r_c}

\def\K{{\rm\thinspace K}}
\def\m{{\rm\thinspace m}}

\def\spose#1{\hbox to 0pt{#1\hss}}
\def\approxlt{\mathrel{\spose{\lower 3pt\hbox{$\sim$}}
	\raise 2.0pt\hbox{$<$}}}
\def\approxgt{\mathrel{\spose{\lower 3pt\hbox{$\sim$}}
	\raise 2.0pt\hbox{$>$}}}

\def\<{\thinspace}
\def\s{\hbox{\phantom{5}}}	

\def\boxit#1{\vbox{\hrule\hbox{\vrule\kern3pt\vbox{\kern3pt
          #1 \kern3pt}\kern3pt\vrule}\hrule}}

\begin{document}

\vskip 1truein
\Large
\centerline{\bf Mergers of Systems Containing Gas}
\normalsize
\vskip 1truein
\centerline{F. R. Pearce$^1$, P. A. Thomas$^1$ \& H. M. P. Couchman$^2$}
\vskip 1truecm
\noindent {\it $^1$ Astronomy Centre, Sussex University, Falmer, Brighton,
BN1 9QH.}

\noindent {\it $^2$ Department of Astronomy, University of Western Ontario,
London, Ontario N6A 3KV, Canada.}

\vskip 1truecm
{\narrower
\noindent{\bf ABSTRACT}

Several simple mergers between model galaxy clusters containing a
mixture of gas and dark matter are examined, testing the coupling of
the gas to the underlying collisionless material.  The gas is shocked,
irreversibly dissipating the energy fed into it by the collisionless
component and forms a resolved constant-density core.  For the dark
matter, however, admixture of phase space vacuum is not very efficient
and a constant-density core is not produced.  In the final state the
central gas has little residual kinetic energy, indicating that
streaming motions do not help to support the gas.

\vskip 1truecm

\noindent{\bf Key Words:} Methods: numerical, Galaxies: clustering,
Galaxies: formation, Galaxies: intergalactic medium, dark matter.
}
\section{Introduction}

In this paper we examine the behaviour of the intra-cluster medium
in galaxy clusters. Cluster gas is important
because it is the dominant form of baryonic matter in these
objects and observations of its X-ray emission provide some of the best
currently available constraints on cluster properties. The gas has
temperatures in the region of $5\times 10^7 \K$ and emits mainly
via thermal \brem. This emission is observed in the X-ray waveband,
where galaxy clusters form some of the brightest sources. As \brem\
emission depends upon the {\it square} of the gas density it is very
sensitive to this parameter and so the surface brightness is sharply
peaked towards the central regions. Much debate has centred around
the observation of a constant density central region in most
clusters (Jones \& Forman 1984) and the relevance of this when
choosing a profile for the underlying dark matter.
We examine the merging scenario to show that constant density gas
cores form naturally {\it without} a similar core in the dark matter.
The question of whether the gas attains the virial temperature is also
investigated. Previous simulations (Evrard 1990, hereafter \Evr,
Thomas \& Couchman 1992, hereafter \TC) suggested that it may not - with
partial support being provided by residual streaming motions. We find no
evidence for this and suggest the motions were due to inadequate resolution
and/or inaccuracies caused by using too long a timestep.

Early attempts to study the behaviour of gas in an astrophysical
context were pioneered by Larson (1978) and extended by Cavaliere
\etal (1986), who used a hydrostatic approximation to model the gas.
This is possible since the high temperatures imply high sound speeds,
which should allow the gas to rapidly adapt to any global changes in
the potential. More recently, attempts have been made to treat the gas
more physically by letting it convert into stars when a certain
condition related to the local density and temperature is reached
(Carlberg 1986).

The development of more sophisticated gas codes (such as \SPH),
that were able
to handle large density contrasts, has enabled several groups to carry out
large simulations modelling cluster scale sections of the universe
(\Evr, \TC, Hernquist \& Katz, 1989, Cen \etal 1990, amongst others).
Usually the largest object in the simulated volume is identified
with a galaxy cluster and studied in detail. The formation history
of this object is often a confused and poorly understood affair
because of the complicated initial state.
As the authors themselves
point out, the resulting structure is not very reliable as the object
concerned contains too few (less than a few thousand) particles. More
recently Navarro \& White (1993) have looked in more detail at simple
mergers but lack the resolution of the simulation described here.

We intend to adopt the approach of simplifying the problem until we
can hope to understand the physics. To achieve this we abandon all
pretence of modelling a definite cosmology, considering objects
similar to those studied in detail in Pearce, Thomas \& Couchman 1993,
(hereafter \PTC), where purely collisionless systems were examined.
We build upon the ground-work laid there by introducing a gaseous
component that makes up 11 percent of the cluster by mass.
The exact amount is not important although observed clusters do contain
about this much gas (See Sarazin 1986 for a general review of
cluster properties, Fabian 1991). By comparing simulations both
with and without a gas component we hope to gain an insight into
the physics of the combined system.

Including gas
in a simulation allows a more direct comparison with observed
phenomena. This is because in the real world the dark matter
distribution must be derived from the visible components (\ie gas and
stars), a process that requires additional assumptions. As our
model contains both phases we can hope to check the accuracy of this
reasoning.

As in \PTC, we try to
understand the merging process in detail.
To do this we deliberately look only at simple mergers,
starting with well understood initial states. Coupled with this
we also have over 32000
particles in our final object, many more than some earlier
simulations. This allows shocks to develop properly, converting the
correct amount of kinetic energy into heat and making the results
obtained from the gaseous component much more reliable.

We collide two sub-clusters, each made up of a mixture of
collisionless (hereafter dark matter) particles and gas.  The ratio,
$\beta$, between the kinetic energy of the dark matter and the thermal
energy of the gas was set equal to unity initially. Then,
\begin{equation}
\beta={\sigma^2 \over {k_BT \over \mu m_H}}
\end{equation}
where $\sigma$ is the one-dimensional velocity dispersion of the
dark matter, $T$ is the gas temperature and $k_B, \mu$ and $\m_H$ are
Boltzmann's Constant, the relative molecular mass
and the mass of a hydrogen atom respectively.

In all our runs we merge two similarly sized objects head-on at
different speeds (off-centre mergers will be considered in a future paper
(Pearce, Thomas \& Couchman 1994)).
We show that the gas is shocked, especially in the centre of the cluster
where its entropy is increased above that of the equivalent measure in
the dark matter. No gas remains with a low
enough entropy to sit at the bottom of the potential well formed by
the dark matter and so a core region of constant density, nearly
constant temperature gas forms. This region is supported almost
entirely by thermal pressure, with little residual kinetic motion and
is in hydrostatic equilibrium. In this region the gas is hotter than
the dark matter, producing values of $\beta \sim 0.8$. (A value of
$\beta < 1$ is consistent with the gas being the more extended
component.)  Beyond the gas core the two phases have very similar
profiles that are close to those obtained in our earlier work (\PTC).
This would indicate that the gas can closely follow the dark matter
profile in the outer regions.

In this paper we are particularly interested in the central regions of the
merger product.
A constant density core region never forms in the dark matter,
confirming the results of \PTC.
For a purely gaseous collision a small core region forms, due to the
shock of dissipating the collisional kinetic energy. In the combination
gas and dark matter runs, however, a larger core forms because in addition
to the initial shock, energy is
exchanged between the two phases whilst they are separated,
and this becomes irreversibly tied to the gas
when it is shocked again by a secondary merger.
It is the combination of energy feed into
the gas and the irreversible jump in entropy
that increases the size of the core region visible in the gas.

In the following section we explain the computational methods used in more
detail and give the specific parameters and configuration of each of our 5
mergers. In Section~3 we present the results, with Section~4
containing
a discussion of them. We present our conclusions in Section~5.

\section{Method}

\subsection{Computational Algorithm}

The majority of the simulations in this paper were carried
out using a combination of smoothed particle hydrodynamics (\SPH) to
follow the gas and
an adaptive particle-particle, particle-mesh (\AP3M) technique to
handle the gravity. We call this hybrid code \APS.

\APS\ is the version of Hugh Couchman's \AP3M\ (Couchman 1991) used by
\PTC\ with gas forces applied at the particle-particle level.
\SPH\ was introduced by Lucy (1977) and Gingold \& Monaghan (1977)
who used it to model polytropic stars.

We use an \SPH\ implementation similar to that of Monaghan, for which the
formalism has been published several times (Gingold \& Monaghan 1982,
Hernquist \& Katz 1989, Monaghan 1992).
When merging \SPH\ and \AP3M\ the
only difficulty arises when a particles smoothing kernel extends over a
subgrid boundary (subgrids are positioned by the \AP3M\
algorithm so that more mesh points are placed in the denser regions.
This reduces the number of
particle-particle calculations that need to be carried out).
We handle this case by checking if a particle has a
neighbour outside the subgrid. If it does then the \SPH\ forces are applied
at the coarser level. A full description and details of the tests
can be found in Pearce (1992) and Couchman, Thomas \& Pearce (1993).
Some general remarks concerning the implementation of \SPH\ can be found
in Martin, Pearce \& Thomas (1993),
hereafter \MPT.

\subsection{Initial Conditions}

We carried out 5 main runs, the parameters for which are listed in Table~1.
All involved colliding two equal mass clusters
at different speeds. Each merger was begun with the two spherically
symmetric clusters just in contact.
The collision speed of run \fs\ was set equal to the circular speed at the
edge of one of the clusters. This is approximately half the one-dimensional
virial speed at the centre of the cluster.
It is also roughly equivalent to the sound
speed in the outer regions. This run is very similar to
run f3 of \PTC, which here we call run \acf\ for consistency of
notation as it contains only dark matter.
We include the parameters in Table~1 for comparison,
as both have the same total mass and collision speed.
The third and fourth runs, \ff\ and \fh\ had collision speeds twice and half
that of run \fs\ respectively. The final run, \go, was carried out at
the same speed as run \fs\ but contained only gas.

The systems we model have initial Hubble density profiles of slope 3;
\begin{equation}
	\rho = \rho_c \left[ 1 + \left({r \over \a}\right) ^2 \right]
                             ^{-{3 \over 2}}
\end{equation}
within some outer radius $\Rmax$, $\rho=0$ beyond $\Rmax$.
Here $\rho_c$ is the central density and $\a$ is the cluster core radius.

The core radius needs to be well within the outer boundary but large
enough so that we can detect a reduction during mergers.  We compromise
on $\Rmax/\a=16$.
The gravitational softening was fixed at $\s=0.4\a$, above the
interparticle separation in the original cores. This is explained
in greater detail in \PTC.

Each phase in each of the initial clusters contained 8192 particles,
giving 32768 particles in all (except for the single component
runs, \acf\ and \go, which had half this number). With each gas
particle having ${1 \over 8}$ of the mass of a dark matter particle
the latter component dominates the overall potential.

The timestep, $\Delta t$, in our simulations is fixed by the condition
\begin{equation}
\Delta t=0.25\min_i\left(d\over a_i\right)^{1\over2},
\end{equation}
where $a_i$ is the acceleration on particle $i$.
$d$ is a distance for which we take;
\begin{equation}
d=\min(\s,h_{min})
\end{equation}
where $h_{min}$ is the minimum smoothing length found by the \SPH\ algorithm
(\ie it is controlled by the greatest density currently in the simulation).
In practice $\s > h_{min}$ at all times for our models, and so we are
applying a more restrictive timestepping criterion than that used in
\PTC. Too long a timestep leads to spurious jumps in a gas particle's
entropy because close interactions are not accurately followed.
This is not a problem in our simulations - see \fig~\ref{statent} and
the discussion in the following section.

The gas is positioned by transforming a relaxed uniform box into the
desired density profile. This ensures that the initial gas separations
are close to a relaxed state immediately. The particles within the
initial box are ordered in radius from an arbitrary centre and then
this radius is adjusted so as to reproduce the density profile. The
local density is set from equation (2), whilst the temperature is set
equal to $\sigma^2$ (\ie $\beta=1$). The velocity dispersion, $\sigma$
is derived by solving Jeans equation for the above density profile
with no outer cut-off (Equation~4-54, Binney \& Tremaine 1987).  The
initial positions of the dark matter are set equal to the gas
positions reflected through the centre of each cluster. In \PTC\ the
initial velocities were set equal to the velocity dispersion at the
particles position. This process only spans five of the six
phase space dimensions. In this paper the velocities
of the dark matter are drawn from a Gaussian
distribution of width $\sigma$, fully spanning the available phase space.
This change makes no discernible difference to the results.
The gas is initially at rest.

\subsection{Tests}

Tests of the individual codes involved have been published elsewhere,
(see \PTC\ for tests of \AP3M, \TC). We tested our version
of \SPH\ by reproducing standard shock results in a 2d by 1d tube.
The correct jump conditions were reproduced with a shock front that was about
6 particles deep. This would indicate that over 1000 particles are required
to model a 3-dimensional shock correctly. Detailed tests of our \SPH\
implementation are given in \MPT\ and \CPT.

A quick and revealing method of looking at global changes in the
system is an energy track. This is illustrated by \fig~\ref{enerfs}
which shows the energy statistics for one of our mergers.  Energy and
time are normalised by setting the total mass, $M_{tot}$, the outer
radius of the system, $R_{max}$ and $G$ equal to unity.  The thermal
energy is displayed multiplied by a factor of 8, as the total gas mass
is ${1 \over 8}$ that of the dark matter.  The effect of the merger is
clearly visible on \fig~\ref{enerfs}, however it is impossible to
judge the accuracy of the energy conservation due to the scaling.
\fig~\ref{enercons} expands the system total energy and scales it to
the initial value.  There are nearly
1000 timesteps in this plot with typical energy conservation better
than 1 part in 10000 per step.  The same accuracy, smoothing and
timestepping parameters were used in all of the runs detailed in this
paper.

\begin{figure}
 \centering
 \caption{Energy conservation}
 \label{enercons}
\end{figure}

The combined code was tested by observing the evolution of a static
configuration. A single isolated cluster was set up following the
prescription given above. This consisted of 8192 gas and 8192 dark
matter particles in what should be a nearly stable arrangement. For
this run the total energy is conserved to better the 0.4 percent and
the start and finish parameters are listed in Table~1.  Some expansion
at the edge is to be expected as here the gas is not in pressure
balance.  As in our previous paper (\PTC) the choice of centre is very
important when plotting logarithmic density profiles. For the static
run we tried the `true' centre (the position about which the initial
profile was built), the position of the gas particle with the highest
local density, the mean position of the 100 or so densest gas
particles, the most tightly bound dark matter particle and the mean of
the 100 or so most tightly bound particles. All five of these
positions remained close together, within less than the gravitational
softening length. In what follows the mean position of the 100 or so
densest gas particles is used as the expansion centre for profiling
and radial measurements.

\begin{figure*}
 \centering
 \caption{Evolution of the gas and dark matter density profiles.}
 \label{densprof}
\end{figure*}

\fig~\ref{densprof} shows the initial and final density profiles
for the gas and the dark matter, with the end-state taken after
approximately the same length of time as the collisions discussed below.
The expansion of the edge can readily be seen, as can a very small
increase in the core size of both the gas and the dark matter.
Some rearrangement in the centre is
almost certain to occur as the existence of a gravitational softening length
is not taken into account when setting up the profiles. This effect will
be most noticeable in the densest regions, particularly so for the gas as
the same particles remain in the centre, whilst the dark matter particles
in the region keep changing. Some small
rearrangement is inevitable anyway as
this distribution function is not formally static.
The small changes in the
core parameters occur quickly and then stop with the system rapidly
settling into a steady state almost indistinguishable from the initial
conditions.

\begin{figure}
 \centering
 \caption{Evolution of the gas temperature and the range in temperatures
at the endpoint.}
 \label{statgastemp}
\end{figure}

\begin{figure}
 \centering
 \caption{Entropy change in a static model.}
 \label{statent}
\end{figure}

The evolution of the gas temperature profile is shown in
\fig~\ref{statgastemp}
along with
the maximum and minimum temperature of particles in each radial bin
at the endpoint. Some cooling has
taken place in the outer regions due to the expansion. This has occurred almost
isentropically. A plot of entropy change
against initial radius is given in \fig~\ref{statent}.
The entropy change, $\Delta S$
is defined as;
\begin{equation}
\Delta S=\log_{10}\left[{T_b \rho_a^{2 \over 3} \over T_a \rho_b^{2 \over 3}}
\right]
\end{equation}
where $T$ and $\rho$ are the temperature and density of a particle and the
subscripts $a$ and $b$ refer to the start and finish times respectively.

The gas has a tendency to order itself in entropy, with low entropy
gas sinking to the bottom of the potential well whilst high entropy gas floats.
It is therefore vitally important to keep the scatter in entropy produced
by the algorithm (the width in \fig~\ref{statent}) as small
as possible, otherwise artificial cool core regions can arise.

The slight upward drift in the mean of \fig~\ref{statent} is caused by a
variety of
reasons. Firstly, the density is overestimated at the first step because
the initial configuration is not completely relaxed, which leads to an
underestimate of the initial entropy. As the
gas relaxes back to a stable configuration the \SPH\ density returns to the
`correct' value. This relaxation is artificial and not due to any actual
density change and so a
slight jump in entropy occurs. Secondly, any application
of the artificial viscosity, which contains sub-sonic terms will increase
the entropy of the particles, so we expect a slow upward drift.
The jumps in a gas particles' entropy due to collisional shocks are over
an order of magnitude greater than these effects.

\begin{table*}
 \caption{Run Parameters, by column a run id, the number of particles in
each phase, the ratio of the collision speed to the central
velocity dispersion, best fit core radii (in units of the gravitational
softening length) and the best fit slopes.}
 \centering
\begin{tabular}{lccccccc}
Label & $N_{DM}$ & $N_{gas}$ & $V_{coll}/\sigma_0$ & $a_{DM}$
& $a_{gas}$ & $s_{DM}$ & $s_{gas}$\\
\fh & 2$\times$8192 & 2$\times$8192 & 0.25 & 2.20 & 3.46 & 2.80 & 2.96\\
\fs & 2$\times$8192 & 2$\times$8192 & 0.5 & 2.36 & 3.51 & 3.08 & 3.01 \\
\ff & 2$\times$8192 & 2$\times$8192 & 1.0 & 2.10 & 4.56 & 2.82 & 2.81 \\
\go & 0 & 2$\times$8192 & 0.5 & - & 3.27 & - & 3.18 \\
\acf & 2$\times$8192 & 0 & 0.5 & 2.22 & - & 2.82 & - \\
single$_{end}$ & 8192 & 8192 & 0 & 2.85 & 3.01 & 3.36 & 3.24 \\
single$_{init}$ & 8192 & 8192 & 0 & 2.50 & 2.50 & 3.00 & 3.00 \\
\end{tabular}
\end{table*}

\section{Results}

\subsection{Dark Matter Only}

The merger of purely collisionless systems has been discussed in
detail in \PTC. We reiterate the general results
here for completeness.

During the merger of two systems of collisionless material the two
dark matter spheres approach, overlap and then the core
regions clearly reseparate before turning around and falling back in
again. The amplitude of the core separation rapidly decreases with
two `bounces' being clearly visible on an energy plot.

The final density profile is close to $\rho \propto r^{-3}$ in the
outer regions and the size of the original constant density core
region drops by almost the maximum amount allowable by phase
space constraints.
This would indicate that no large scale constant density core region
should exist in the dark matter (and potential well and mass) profiles
of observable objects. (Note, however, that these models ignore the influence
of a central
galaxy, which may sweep the core free of dark matter via two-body
interactions.)

\begin{figure*}
 \centering
 \caption{Potential Energy for runs \acf, \go\ and \fs, scaled
as in \fig~9.}
 \label{enerplot}
\end{figure*}

In \fig~\ref{enerplot} we give the potential energy tracks
for this run (\acf), the
pure gas run (\go) and the combination gas and dark matter run carried out
at this collision speed (\fs).

The final density profile for both this pure dark matter run and the pure
gas run (\go) mentioned below are shown in \fig~\ref{f3gas}

\begin{figure*}
 \centering
 \caption{Density profiles from runs \acf\ and \go.
The densities are scaled so that the initial central density
is unity.}
 \label{f3gas}
\end{figure*}

\subsection{Gas Only}

The merger of two purely gaseous spheres is in effect the simplest of all
the 5 runs considered here. During the collision the forward side of
each of the spheres
is compressed forming an enhanced density slab perpendicular to the
collision axis. The isotropic nature of the gas forces rapidly destroy
this structure in favour of a spherically symmetric density profile.
In the final state the original gas spheres
form two hemispheres aligned along the
collision axis with little interpenetration occurring. Only one
deepening of the potential is seen in \fig~\ref{enerplot} and there are no
post-merger oscillations.

The speed with which a spherically symmetric structure is obtained is
quite surprising, with the density enhancement along the collision
front being smoothed out over a period close to that for local sound
crossing.  The only sign of a recent merger event is a temperature
enhancement and two approximately planar shock fronts propagating
outwards perpendicular to the collision axis.

The final static structure has the density profile shown in
\fig~\ref{f3gas}.
This is close to
the purely dark matter case, with only a small constant density
core region forming and a slope close to $\rho \propto r^{-3}$
in the outer regions.

To aid in analysing the merger we binned the gas particles into 4 groups
depending upon their initial position.
The inner 25 percent by mass we refer to as the core gas, whilst
the outer 25 percent forms the halo gas. There are also two intermediate bins
comprised of the gas between 25 percent and 50 percent ($1/2$ gas) and
between 50 percent and 75 percent ($3/4$ gas). Each bin contains $4000$ gas
particles with the outer $192$ from each initial cluster being ignored
for this part of the analysis.
There is little mixing between the four groups, with original halo
gas remaining, on the whole, as halo gas and original core gas
remaining as core gas.

\begin{figure*}
 \centering
 \caption{Entropy, Kinetic energy and total energy of the core gas. The
entropy is scaled arbitrarily whilst the energies are scaled as in \fig~9}
 \label{gasbin}
\end{figure*}

In \fig~\ref{gasbin} we show how the kinetic and total energy of the core
gas varies with time.
The only significant feature appears as the two cores
merge. The kinetic energy picked up during the approach
is thermalised leaving the gas almost stationary. This sudden stop
is also seen in the entropy curve plotted above (and in \fig~\ref{gasentbin}).
The core gas loses energy during the collision because it does work on the gas
immediately surrounding it.
A shock wave propagates outwards,
progressively doing work on gas further and further out. The halo gas is
given a significant amount of energy by this process,
becoming unbound in our case.

\begin{figure*}
 \centering
 \caption{Entropy track for the gas binned by initial radius.}
 \label{gasentbin}
\end{figure*}

The gas is shocked by the collision, with a small constant density
core forming.
\fig~\ref{gasentbin} shows the entropy of each of the 4
gas bins plotted against time and scaled to their initial values. The
shock of the collision can clearly be seen along with the large boost
in entropy which the halo gas undergoes as it is shocked by
gas interior to it. The entropy jump is smallest for the
core gas in this case. This will contrast with the mixed runs which we
discuss next.

\subsection{Dark Matter and Gas}

We have carried out 3 combination dark matter and gas runs. The main
one had the same collision speed as the pure dark matter and pure gas
runs discussed above. This run was bracketed by runs at half and twice
this speed. The slowest run corresponds to two clusters forming almost
in contact whilst the fastest run is only bound because of the internal
energy of the clusters. The central speed would be obtained by two point
masses starting from rest at an initial separation of 4 times the maximum
radius of one of the clusters.
In all three cases the general scenario is the same. The
merger follows a path similar to the purely collisionless case until
maximum compression is reached. The two gas spheres collide and form a
pancake like structure about the centre of mass of the system, with
the short axis aligned with the collision direction. The dark matter
component continues on, leaving its original gas behind. At this point
the gas is at rest around the centre of mass of the system, whilst the
collisionless material has re-separated as in the dark matter only
case discussed above. The two dark matter centres are clearly distinct
and some distance from the centre of mass.  The central gas
is no longer confined in a deep potential well so it expands, falling
back towards the nearest receding dark matter sphere (which in
the head-on case is not the one it started from).

\begin{figure*}
 \centering
 \caption{Energy plot from the \fs\ merger. The energy is
in units of the total initial mass, the maximum radius of one of the
initial clusters ($\Rmax$) and $G=1$.
The time unit is taken to be $(\Rmax^3/GM)^{1 \over 2}$.}
 \label{enerfs}
\end{figure*}

As with the collisionless mergers the simplest way to follow the large scale
evolution of the system is an energy plot. \fig~\ref{enerfs}
shows the total energy, the total kinetic energy and eight times
the thermal energy of the gas for run \fs. Two passages are clearly visible,
although only the first of these has a significant impact on the total thermal
energy of the gas. The system was allowed to settle
down for several crossing times before the
steady state properties of the final object were examined.

A comparison of the potential energy plots for this run, a similar run
without gas (\acf) and a purely gaseous run (\go), is given in
\fig~\ref{enerplot}.
The presence of even a small amount of
gas causes faster evolution towards the final state after the initial
merger event has taken place. This is because each
cluster deposits its gas at the centre of mass on the first collision
and then this
mass drags back the receding dark matter components. The time gap between
the first and second passage through the centre of mass is significantly
reduced in this way. While the gaseous and collisionless components are
well separated energy is exchanged between them.

\begin{figure*}
 \centering
 \caption{Density profiles for the gas and dark matter at the end of
run \fs.
The densities are scaled so that the initial central density
is unity. The gas density has been multiplied by 8, scaling it up to match
the dark matter.}
 \label{denscom}
\end{figure*}

The density profiles for the gas and dark matter at the end of run \fs\,
are given in \fig~\ref{denscom}. The
gas is in hydrostatic equilibrium, and the profiles have stopped evolving.
In this combination run the two phases no longer have
the same profile, with
a definite constant density core appearing in the gas.
Contrast this evolution with the
static model of \fig~\ref{densprof},
where the two phases stay closely linked.

\begin{figure*}
 \centering
 \caption{Entropy, Kinetic energy and total energy of the core gas from
the combined run. Again the
entropy is scaled arbitrarily whilst the energies are scaled as in \fig~9}
 \label{gasdmbin}
\end{figure*}

Again, as for the purely gaseous case we bin the gas depending upon
its position in the original clusters.
Initially similar behaviour is observed,
however comparison of \fig~\ref{gasdmbin} with \fig~\ref{gasbin} reveals that
extra structure appears after the first merger.
Once the core gas has been halted in a shock
and undergone an entropy jump, energy is pumped
into it from the receding dark matter.
The gas is not confined at the centre of the system and so expands, falling
back into the dark matter potential wells. The net effect is to transfer
the orbital energy of the dark matter in kinetic energy of the gas.
The dark matter spheres
do not reach the same separation as in the purely collisionless state,
clearly showing that some orbital energy has been lost to the gas.

\begin{figure*}
 \centering
 \caption{Entropy track for the gas in the combined run binned by
initial radius.}
 \label{gasdmentbin}
\end{figure*}

The main additional feature on \fig~\ref{gasdmbin}
that is not on \fig~\ref{gasbin} is the boost in entropy experienced
by the core gas during a second collision.
Upon turn around and re-collapse the core gas, which has gained
energy from the dark matter, is heavily shocked, irreversibly
increasing its entropy. This second shock
is clearly visible in \fig~\ref{gasdmentbin}, where the
entropy of the gas in run \fs\ is plotted against time (again relative
to the initial values).
Only the core gas experiences this strong
second shock as only it undergoes a second collision.
The entropy of the core gas is raised to such an extent that no low
entropy gas remains that is capable of sitting at the bottom of the
potential well now formed by the dark matter.  In this case the core
gas has been more heavily shocked than the gas surrounding it but remains
at lower entropy.

\begin{figure}
 \centering
 \caption{Temperature profile at the end of run \fs.}
 \label{gastemp}
\end{figure}

The final temperature profile is given in \fig~\ref{gastemp}.
The small spread
between the maximum and minimum temperatures in any bin is an indication
of the reliability of the results. (Spurious cooling, an artifact of
misconceptions in the application of \SPH, can lead to cold particles which
sink because of their low entropy, producing an artificial cool region
at the centre. The existence of these
particles is masked when only the mean temperature is plotted.) The temperature
profile is nearly isothermal in the central region, slowly rolling off beyond
the gas core radius. We find no evidence for a central temperature inversion
found by previous authors (\Evr,\TC).

The gas at the end of our simulation has very little residual kinetic
energy. The ratio of core gas kinetic to thermal energy is about 4 percent
imediately after the merger, dropping to
1.65 percent by the end due to the viscous terms in the code.
These values are lower than those presented by previous studies (\Evr,\TC).
We believe the difference is due (as suggested by them) to incomplete
thermalisation of the kinetic
energy due to the small number of gas particles in
the interesting regions of the earlier work. Our results contain no sign
of large scale streaming motions in the gas, suggesting that these motions
are not responsible for partial support of the gas in real clusters
(However, preliminary results from offset mergers suggest that rotational
kinetic energy can amount to as much as 30 percent of the thermal energy).

The results for both the faster and slower collisions, runs \ff\ and \fh\
are very similar to those presented
above. In both the dark matter core size drops whilst a constant
density gas core appears.
The final gas core is larger in run \ff, whilst for
run \fh\ it is smaller. This is to be expected as during faster mergers
the initial shock is greater and the gas picks up more
kinetic
energy before the second collsion of the cores.
Table~1 lists the measured core sizes
for both the dark matter and gas phases for all 5 of the runs presented
above. In all cases the fit was carried out between a radius of 2 and 30
gravitational softening lengths with a Hubble density profile of the form:
\begin{equation}
	\rho = \rho_c \left[ 1 + \left({r \over \a}\right) ^2 \right]
                             ^{-{s \over 2}}
\end{equation}
which is similar to equation~(2) except that we now have a variable
slope, $s$,
(instead of fixing $s=3$). The best fit value for $s$ is also
listed in Table~1.

The existence of a gas core radius
{\it does not} reveal a similar dark matter core, whereas at larger radii
the gas and dark matter do appear to have very similar profiles.
\section{Discussion}

\subsection{Real Clusters}

X-ray observations of galaxy clusters have provided evidence for a
core region in the gas (Jones and Forman 1984), but the interpretation
of this result was complicated by the large beam size of the detectors
used (see Gerbal \etal 1992).  It is often assumed that the underlying
gravitational potential has a core radius as the gas sound speed is
high enough to allow many crossing times in the central regions. We
find that the two distributions need not be so closely related, with
the central gas being shocked more than the equivalent process of
phase mixing for the dark matter during our collisions. The gas can
then exist in hydrostatic equilibrium in the core-less potential well
defined by the dark matter whilst maintaining a constant density core
region itself.

The problem of comparing our models to the observations is complicated by the
lack of a galactic population in the simulations. Although we can observe the
gas via its X-ray emission,
in the real world the dark matter distribution must be derived
from that of the galaxies. Processes such as dynamical friction and
galaxy-galaxy mergers will drive these two distribution functions apart,
particularly in the dense central regions. In addition the presence of a large
central galaxy in many clusters is invariably associated with a cooling flow
and may also alter the core dark matter distribution via two-body interactions.

Coupled with this the formation of real galaxy clusters is likely to be a lot
more messy than our simple mechanism, involving multi-body mergers with a range
of masses, speeds and impact parameters. The colliding clusters may also
contain a significant amount of sub-structure, with secondary infall and tidal
torques providing additional complications.  Preliminary results for offset
mergers, to be reported in a future paper, show that a gas core is obtained for
all impact parameters, formed via the same shock mechanism detailed here plus
some additional rotational support.

\subsection{The relative energies of the gas and dark matter}

In the cores of the merger products in our simulations the gas is hotter than
the dark matter, $\beta<1$ (see Equation~1).  However outside the core,
throughout the majority of the cluster, we have $\beta\approx1$.  This latter
result is in
agreement with previous simulations (\Evr,\TC).  As a consequence of this
equality of specific energies for the gas and dark matter, the spatial
distribution of the two phases are very similar.  However the actual form of
the density profile varies in different simulations.  Merger simulations ({\it
e.g.} White 1980, Villumsen 1983, \PTC, this work) tend to give slopes at the
half-mass radius of $s\approx3$.  More realistic simulations with continuous
infall of matter give shallower slopes, $s\approx2$ (\Evr,\TC).

The amount of kinetic energy present in the inner third of the gas at the
endpoints of simulation \fs\ is just 1.6 percent of the total thermal plus
kinetic energy.  Most of the thermalisation of the kinetic energy takes place
in the two major shocking events associated with the collisions of the gas
cores.  However there is a slow downward drift of the kinetic energy at later
times---immediately after the second collision the kinetic energy is 4 percent
of the total.  It is not clear to what extent this continued thermalisation is
to be expected in real systems as the dissipation processes within the
intracluster medium are not well-understood.  Nevertheless, the low fractional
kinetic energies which we find are smaller than those recorded by \Evr\ and
\TC.  There are two possible reasons for this.  Firstly, it could be that our
integrations are more accurate and that the motions seen in previous work were
numerical artefacts caused by having too few particles in the cluster.
Alternatively, the head-on collisions reported here might be singular.
Preliminary results from off-centre collisions suggest that as much as 30
percent of the energy of the gas may be in the form of rotational kinetic
energy.

\subsection{Adding low entropy gas}

One way of eliminating the finite gas core in our simulations would be to add
low entropy gas to the cluster.  This might be, for example, by the infall of
cool subgroups, or by the expulsion of cold gas from galaxies.  The difficulty
with this is that the speeds of such systems relative to the cluster centre
would in general be larger than the sound speed and shocks would raise the
entropy up to that of the cluster.  For example the slow collision considered
in this paper had an initial approach speed that was equivalent to starting the
clusters from rest with their centres just 2.3 cluster radii apart, and yet the
core was almost as large as for the more violent collisions.

\subsection{Disclaimer}

We should stress that as yet no simulation of the intracluster medium has
attempted to model the physical processes which must be going on within it.
We know that the metallicity of the gas is about 0.3 solar and this influx of
metals is associated with an unknown injection of energy.  Gas and magnetic
fields are continually being stripped from infalling galaxies and there is an
exchange of energy between these galaxies and the gas and dark matter.  We have
deliberately kept our simulations simple so that the collision process can be
examined in isolation.  Our clusters may not, however, resemble those in the
real world.

\section{Conclusions}

In \PTC\ we showed that successive collisions tend to reduce the core radii of
the collisionless component as little phase vacuum is mixed into the central
regions during mergers. This result holds true for the dark matter in the
dual-component mergers studied in this paper.

The gas-only merger products show a small increase in core radius due to
shocking of the central gas and this effect is exaggerated in the two-phase
collisions.  The specific energy of the core gas is higher than that of the
dark matter and so the former is more extended.  At larger radii the two
components have very similar profiles, close to the \dev\ profile found by
\PTC\ (which is also approximately their initial profile). Secondary infall
may lower the slope of the density profile in real clusters (Bertschinger
1985).

The lack of any residual kinetic energy in the gas in the central part of our
simulation, in contrast to previous work, indicates that many gas particles are
needed in order to model shocks correctly.  Earlier studies have contained
around 1000 gas particles in the interesting regions whilst our models contain
over 16000.  Structures containing less than a few thousand gas particles that
have been modelled with \SPH\ and have formed via shocking {\it cannot} be
trusted in detail. This is an important constraint on cosmological simulations,
which although they often contain many more particles {\it in total} than we
are capable of following, also have far more elaborate structures and many more
condensations each of which consists of relatively few particles.

How generic are the results presented in this paper?  We have suggested that
energy is pumped from the dark matter into the gas and then irreversibly
dissipated in shocks.  Preliminary simulations involving off-centre collisions
suggest that a similar mechanism operates in more complicated situations.
Additional support of the core is provided by rotation.  Our idealised
models would therefore suggest a constant density gas core region in all
clusters, with a correspondingly small core radius for the underlying dark
matter. Unfortunately this conclusion neglects a wide range of physical
processes as discussed above.

We believe that the combination of an adaptive high resolution particle code
and smoothed particle hydrodynamics, together with our approach of looking at
relatively simple systems has allowed us to reach new and firm conclusions
about the properties of the intracluster gas in galaxy clusters and the
underlying dark matter distribution.

\section*{Acknowledgements}

We would like to thank Simon White for a discussion on the physical
mechanism of core production.  PAT and HMPC acknowledge a NATO
Collaborative Research Grant CR6920182 which facilitated their
interaction.  We are grateful for use of the facilities of the
STARLINK minor node at Sussex.

\section*{References}
\parindent=0pt\parskip=4pt\hangindent=3pc\hangafter=1

Bertschinger, E., 1985, \ApJS , 58, 39

Binney, J., Tremaine, S., 1987, Galactic Dynamics.
Princeton University Press, Princeton

Cavaliere, A., Santangelo, P., Tarquini, G., Vittorio, N.,
1986, \ApJ ,305, 651

Carlberg, R., 1986, \ApJ , 310, 593

Cen, R. Y., Jameson, A., Lin, F., Ostriker, J. P., 1990, \ApJL , 326, L41

Couchman, H. M. P., 1991, \ApJL , 368, L23

Couchman, H. M. P., Thomas, P. A., Pearce, F. R., 1993, \prep, \CPT

Evrard, A. E., 1990, \ApJ , 363, 349, \Evr

Fabian, A. C., 1991, \MN , 253, 29p

Gerbal, D., Durret, F., Lima-Neto, G., Lachiezo-Rey, M., 1992,
\AaA , 253, 77

Gingold, R. A., Monaghan, J. J., 1977, \MN , 181, 375

Gingold, R. A., Monaghan, J. J., 1982, \JCP , 46, 429

Hernquist, L., Katz, N., 1989, \ApJS , 70, 419

Jones, C., Forman, W., 1984, \ApJ , 276, 38

Larson, R. B., 1978, \MN , 184, 69

Lucy, L. B., 1977, \AJ , 82, 1013

Martin, T. J., Pearce, F. R., Thomas, P. A., 1993, \Pre, University of Sussex,
\MPT

Monaghan, J. J., 1992, \ARAA , 30, 543

Monaghan, J. J., Lattanzio, J. C., 1985, \AaA , 149, 135

Navarro, J. F., White, S. D. M., 1993, \Pre , University of Durham

Pearce, F. R., 1992, \PhD, University of Sussex

Pearce, F. R., Thomas, P. A., Couchman, H. M. P., 1993, \MN , 264, 497, \PTC

Pearce, F. R., Thomas, P. A., Couchman, H. M. P., 1994, \Pre

Sarazin, C. L., 1986, \RMP , 58, No.1

Thomas, P. A. , Couchman, H. M. P., 1992, \MN , 257, 11, \TC

Thomas, P. A. , Fabian, A. C., 1992, \MN , 246, 156

Villumsen, J. V., 1983, \MN , 204, 219

White, S. D. M., 1980, \MN , 191, 1p

\end{document}